%% file: main.tex
\documentclass[runningheads]{llncs}
\usepackage[T1]{fontenc}
\input{packages.tex}

\input{command.tex}

\let\llncssubparagraph\subparagraph
\let\subparagraph\paragraph
\usepackage{titlesec}

\titleformat{\subsection}[runin]
  {\normalfont\bfseries}    
  {\thesubsection}          
  {1em}                     
  {}                        

\titlespacing*{\subsection}
  {0pt}                     
  {1.5ex plus 1ex minus .2ex} 
  {1em}                     
\let\subparagraph\llncssubparagraph

\begin{document}
\title{LLMATCH: a Unified Schema Matching Framework with Large Language Models \thanks{\scriptsize{Supported by the Singapore Ministry of Education Academic Research Fund Tier 1 grant (23-SIS-SMU-063) and the SMU-SUTD Internal Research Grant Call (SMU-SUTD 2023\_02\_01)}}}

\titlerunning{LLMATCH: a Unified Schema Matching Framework with LLM}

\author{Sha Wang\inst{1} \and
Yuchen Li\inst{1} \and
Hanhua Xiao\inst{1} \and
Bing Tian Dai\inst{1} \and
Roy Ka-Wei Lee\inst{2} \and
Yanfei Dong\inst{3} \and
Lambert Deng\inst{4}
}

\authorrunning{S. Wang et al.}

\institute{Singapore Management University \\
\email{\{sha.wang.2021, hhxiao.2020\}@phdcs.smu.edu.sg,\\ \{yuchenli, btdai\}@smu.edu.sg}, \and
Singapore University of Technology and Design 
\email{roy\_lee@sutd.edu.sg}, \and
PayPal
\email{dyanfei@paypal.com}, \textsuperscript{4} DBS Bank, \email{lambertdeng@dbs.com}
}

\maketitle        
\begin{abstract}
\import{./sections/}{abstract}

\end{abstract}

\section{Introduction}
\label{chap:introduction}
\import{./sections/}{introduction}

\section{Related Work}
\label{chap:related_work}
\import{./sections/}{related_work}

\section{Problem Definition}
\label{chap:problem_statement}
\import{./sections/}{problem_statement}

\section{\framework Framework}
\label{chap:methodology}

\import{./sections/}{methodology}

\section{Dataset Development}
\import{./sections/}{dataset}
\section{Evaluation}
\label{chap:evaluation}
\import{./sections/}{evaluation}

\section{Conclusion}
\label{chap:conclusion}
\import{./sections/}{conclusion}

\bibliographystyle{splncs04}
\bibliography{main}

\end{document}

%% file: packages.tex
\usepackage{import}
\usepackage{amsmath}
\usepackage{comment}
\usepackage{xspace}
\usepackage{amsmath,amssymb,amsfonts}
\usepackage{algorithm}
\usepackage{algpseudocode}
 \usepackage{multirow} 
\usepackage{textcomp}
\usepackage[smartEllipses]{markdown}
\usepackage{listings}
\usepackage{makecell}
\usepackage{tabularx}
\usepackage{subcaption}
\usepackage{threeparttable}
\usepackage{wrapfig}

\usepackage{enumitem}
\usepackage{amssymb}
\usepackage{graphicx}
\usepackage[dvipsnames]{xcolor}
\usepackage{hyperref}
\hypersetup{
    colorlinks=true,
    linkcolor=blue,
    filecolor=black,      
    urlcolor=cyan,
}
\usepackage[format=default,labelfont=bf]{caption}
\usepackage{changepage}
\usepackage{adjustbox}
\usepackage{array}
\newcolumntype{P}[1]{>{\centering\arraybackslash}p{#1}}
\newcolumntype{R}[1]{>{\raggedleft\arraybackslash}p{#1}}

\usepackage{tikz}
\usetikzlibrary{shapes.geometric, arrows}
\usepackage{booktabs} 

\tikzstyle{process} = [rectangle, rounded corners, minimum width=3cm, minimum height=1cm, text centered, draw=black]
\tikzstyle{preprocess} = [rectangle, dashed, rounded corners, minimum width=3cm, minimum height=1cm, text centered, draw=black]
\tikzstyle{arrow} = [thick,->,>=stealth]

%% file: command.tex
\usepackage{enumitem}

\newcommand{\framework}{\textsc{LLMatch}\xspace}

\newcommand{\benchmark}{\textsc{SchemaNet}\xspace}

\newcommand{\rematch}{\textsc{ReMatch}\xspace}
\newcommand{\unicorn}{\textsc{Unicorn}\xspace}
\newcommand{\coma}{\textsc{Coma}\xspace}
\newcommand{\similarityflooding}{\textsc{SF}\xspace}
\newcommand{\cupid}{\textsc{Cupid}\xspace}

\usepackage{pgfplots}
\usetikzlibrary{pgfplots.groupplots}
\usetikzlibrary{patterns}
\usepackage{xcolor}
\definecolor{lightpurple}{HTML}{c3a2da}
\definecolor{pink}{HTML}{e3adc8}
\definecolor{lightblue}{HTML}{5092c7}
\definecolor{yellow}{HTML}{d4c483}
\definecolor{darkblue}{HTML}{13345e}

\definecolor{darkpurple}{HTML}{9468b1}
\definecolor{darkpink}{HTML}{c17a9b}
\definecolor{darkblue}{HTML}{27689b}
\definecolor{darkyellow}{HTML}{a39053}

\pgfplotsset{
    customBarChartStyle/.style={
        width=\textwidth,
        height=10cm,
        ybar,
        bar width=0.5cm,
        xlabel={},
        ylabel={F1 Score},
        symbolic x coords={IMDB-Sakila, CMS-OMOP, Synthea-OMOP, CPRD Aurum-OMOP, CPRD Gold-OMOP, MIMIC-OMOP},
        xtick=data,
        x tick label style={rotate=9, font=\scriptsize},
        ymin=0, ymax=1,
        legend style={at={(0.8,0.95)}, anchor=north, legend columns=1},
        nodes near coords,
        nodes near coords align={vertical},
        every node near coord/.append style={font=\footnotesize}
    },
    barStyleA/.style={
        draw=darkblue,
        semithick,
        pattern=north east lines,
        pattern color=darkblue,
    },
    barStyleB/.style={
        draw=darkpink,
        pattern=north west lines,
        pattern color=darkpink,
    },
    barStyleC/.style={
        draw=darkpurple,
        semithick,
        pattern=grid,
        pattern color=darkpurple,
    },
    barStyleD/.style={
        draw=darkyellow,
        semithick,
        pattern=crosshatch,
        pattern color=darkyellow,
    }
}

%% file: sections/abstract.tex
Schema matching is a foundational task in enterprise data integration, aiming to align disparate data sources. While traditional methods handle simple one-to-one table mappings, they often struggle with complex multi-table schema matching in real-world applications. We present \framework, a unified and modular schema matching framework. \framework decomposes schema matching into three distinct stages: schema preparation, table-candidate selection, and column-level alignment, enabling component-level evaluation and future-proof compatibility. It includes a novel two-stage optimization strategy: a \textit{Rollup} module that consolidates semantically related columns into higher-order concepts, followed by a \textit{Drilldown} module that re-expands these concepts for fine-grained column mapping. 
To address the scarcity of complex semantic matching benchmarks, we introduce \benchmark, a benchmark derived from real-world schema pairs across three enterprise domains, designed to capture the challenges of multi-table schema alignment in practical settings.
Experiments demonstrate that \framework significantly improves matching accuracy in complex schema matching settings and substantially boosts engineer productivity in real-world data integration.

%% file: sections/introduction.tex

Schema matching is a database system task that identifies semantic correspondences between columns of source and target schemas and transforms them into a unified format. Industries such as healthcare~\cite{reich2024ohdsi}, finance~\cite{iso20022standard}, energy~\cite{pan2023schema}, and agriculture~\cite{asif2023semi} are increasingly adopting standardized data formats to improve interoperability. 
In healthcare, unified data models such as OMOP~\cite{omop} have demonstrated benefits for cross-institutional research and treatment discovery, driving efforts to migrate existing schemas to standardized targets~\cite{overhage2012validation,ohdsi2019book}. However, flawed data migration can have serious consequences, as illustrated by Deutsche Bank’s 2023 account lockouts due to integration failures with Postbank~\cite{iso20022standard,FT_DeutscheBank2024}. Schema matching in real-world scenarios is complex; while research has primarily addressed basic table-to-table alignment, industrial applications often involve mapping multiple source and target tables, requiring a comprehensive understanding of schema semantics, inter-table relationships, and structural dependencies such as primary–foreign keys and hierarchies.

\begin{wrapfigure}{l}{0.4\textwidth}
    \centering
    \vspace{-9mm}\includegraphics[width=0.4\textwidth]{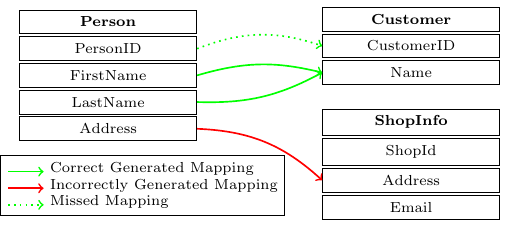}
      \caption{\textit{Toy matching task using traditional methods }}
    \label{fig:schema_matching_example}
    \vspace{-8mm}
\end{wrapfigure}

Before the advent of large language models (LLMs), complex multi-table schema matching was largely intractable~\cite{zhang2023schema}. Traditional tools struggled with challenges such as inconsistent naming conventions, data heterogeneity, and the lack of standardized structural representations. As shown in Fig.~\ref{fig:schema_matching_example}, traditional embedding-based NLP methods are unable to distinguish minute differences among attributes such as \textit{Person.address} and \textit{ShopInfo.address}, leading to incorrect mappings. 
Resolving such mappings often demands extensive, error-prone manual effort by domain experts, with a single schema alignment taking up to 500 hours for two experts.~\cite{ETLLambdaBuilder,paris2021transformation,ackerman2019cognitive,shraga2021learning}.

 The emergence of LLMs has enabled a new class of schema matching methods with strong performance across diverse tasks~\cite{wornow2024automating,10.14778/3611479.3611527,rematch,huang2150transform}. However, two key limitations persist: (1) prevailing approaches treat schema matching as a monolithic process and evaluate it end-to-end, obscuring the contribution of individual components; (2) they largely focus on pairwise table matching, neglecting the more complex many-to-many mappings common in real-world applications.

\begin{figure}[t]
    \centering
    \includegraphics[width=\linewidth]{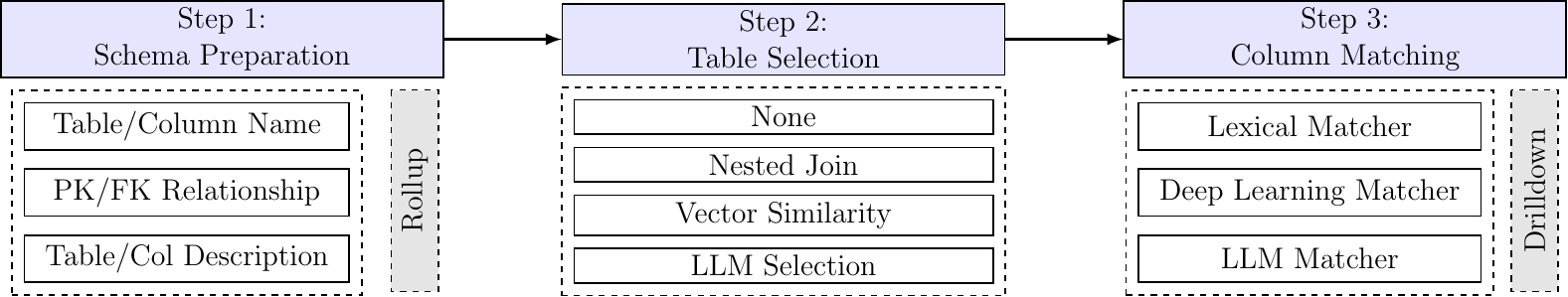}
            \vspace{-6mm}
            \caption{\small Architecture diagram of \framework. \emph{Schema preparation} organises information into three key categories for downstream processing. \emph{Table selection} maps target tables to source tables. \emph{Column matching} performs column-level mapping.}
            \label{fig:flowchart}
             \vspace{-4mm}
\end{figure}

In this paper, we propose a unified, modular framework for schema matching that decomposes the process into three stages: \emph{schema preparation}, \emph{table selection}, and \emph{column matching} (Fig.~\ref{fig:flowchart}). \emph{Schema preparation} organizes the information of source and target schemas into standard categories including names, relationships and metadata. \emph{Table selection} narrows the candidate target tables for each source table, reducing the context length for downstream processing. Finally, \emph{column matching} aligns columns between each source table and its selected target tables to produce the final mappings. This modularization promotes compatibility with existing and future methods while enabling fine-grained analysis of component-level contributions to overall performance.

\begin{figure}[t]
\includegraphics[width=\linewidth]{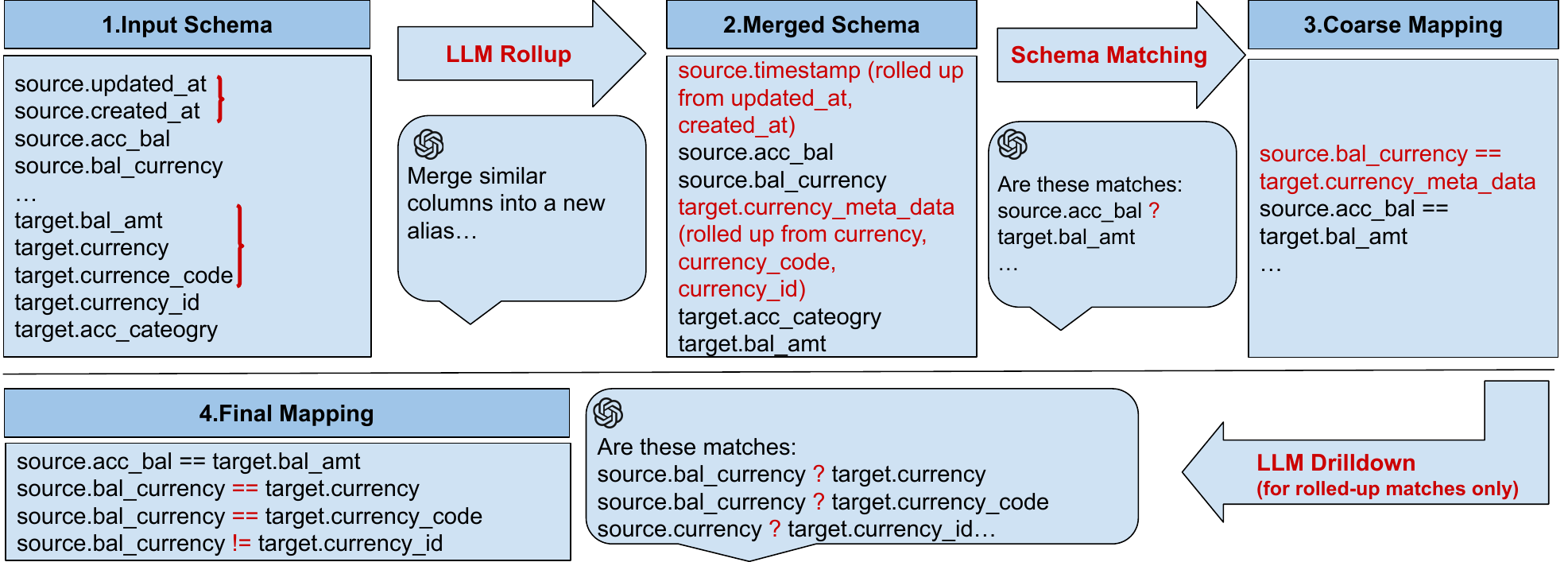}
          \vspace{-4mm}
            \caption{\small Example of multi-level matching with Rollup and Drilldown}
            \label{fig:rollup-drilldown}

    \vspace{-7mm}
\end{figure}

To improve multi-table schema matching, we introduce an optimization strategy based on \emph{Rollup} and \emph{Drilldown} techniques, applied on \emph{schema preparation} and \emph{column matching}, respectively (Fig.~\ref{fig:rollup-drilldown}). \emph{Rollup}  aggregates semantically related columns into a higher-level abstraction. For instance, time-related fields like \texttt{updated\_at} and \texttt{created\_at} are abstracted as \texttt{timestamp}. The abstracted schemas are then used to perform coarse-grained alignment. Upon identifying a high-level match, the \emph{Drilldown} phase reverses the abstraction to perform fine-grained column alignment. This hierarchical approach greatly enhances performance on large and semantically rich schemas while reducing computational overhead and improving scalability.

Recognizing the lack of large-scale, semantically rich benchmarks for schema matching, we curated \benchmark, a comprehensive benchmark comprising seven datasets across the finance, healthcare, and entertainment domains. Developed in collaboration with two leading financial institutions, \benchmark captures real-world data integration challenges and encompasses a wide range of schema matching complexities. On average, each dataset includes 14 tables and 135 columns, substantially exceeding the scope of existing benchmarks, which typically feature a single table with around 15 columns (Table~\ref{tab:groundtruth}). \benchmark offers a more realistic and diverse foundation for evaluating schema matching methods in practical, multi-table settings.


In summary, this work makes the following key contributions:
\begin{itemize}[label=•]
\item We present \framework, a unified and modular schema matching framework that supports existing and future methods while enabling fine-grained evaluation of component contributions.

\item We develop an optimization strategy leveraging \emph{Rollup} and \emph{Drilldown} techniques to address the challenges of complex multi-table schema matching.  

\item We introduce \benchmark, a multi-domain benchmark co-developed with leading financial institutions, designed to capture the complexity of real-world multi-table schema alignment.

\item We conduct extensive experiments demonstrating state-of-the-art performance on 5 benchmarks including \benchmark, with substantial gains on large, semantically rich datasets.  

\item Both \framework and \benchmark are made public to facilitate further research in schema matching~\cite{sourcecode}.  

\end{itemize}

    
    


%% file: sections/related_work.tex
\noindent Schema matching methods generally rely on two types of information: (1) schema metadata and (2) instance data. We briefly review both approaches, followed by recent developments using deep learning and large language models (LLMs). 
\\\textbf{Schema-based Approaches.} These methods leverage metadata such as attribute names, data types, and table relationships. Linguistic features help identify semantic similarities~\cite{do2002coma,cupid,munir2014instance}, while constraints like keys and uniqueness provide structural cues~\cite{zhao2007combining,bernstein2011generic}. More advanced techniques incorporate external resources (e.g., dictionaries, ontologies) to resolve abbreviations and domain-specific terms~\cite{sorrentino2010schema,giunchiglia2004s}. However, their effectiveness diminishes when metadata is sparse or inconsistent.  \\\textbf{Instance-based Approaches.} When metadata is insufficient, instance-based methods analyze data values and distributions to infer correspondences~\cite{munir2014instance,mehdi2017approach}. ML-based matchers learn from patterns in instance data~\cite{rong2012machine}, but often require large training sets and retraining for new domains~\cite{feng2009instance}. These methods also struggle with scalability and may underperform on large, noisy datasets. Hybrid approaches combine schema and instance information to improve accuracy~\cite{asif2023semi,shraga2020adnev}. \\\textbf{Deep Learning and LLM-based Approaches.} Recent methods leverage embeddings and pre-trained models to capture deeper semantic and structural patterns. SMAT~\cite{zhang2021smat}, SemProp~\cite{fernandez2018seeping}, and EmbDI~\cite{cappuzzo2021embdi} use word and instance embeddings, while REMA~\cite{koutras2020rema} applies graph embeddings. LLM-based approaches go further: LSM~\cite{zhang2023schema} integrates active learning; Parciak et al.~\cite{parciak2024schema} use LLMs for single-table matching; and Huang et al.~\cite{huang2150transform} combine offline PK/FK inference with iterative SQL-based refinement. Our work differs by avoiding query generation and human-in-the-loop feedback, and by supporting complex multi-table matching. \rematch~\cite{rematch} shows strong results with GPT-4~\cite{achiam2023gpt}, but performance degrades with smaller models like GPT-3.5~\cite{openai2023gpt35}, which often fails to capture relational structures.

%% file: sections/problem_statement.tex
A database schema \( S \) consists of a set of tables \( T \), columns \( A \), descriptions of tables and columns \( D \), and relationships \( R \), including primary keys (PK) and foreign keys (FK).
Given two database schemas \( S_s \) (source) and \( S_t \) (target), the objective is to map tables and columns from \( S_s \) to those in \( S_t \). 



\noindent\textbf{Motivation for n:m Mapping:} In many real-world data integration scenarios, especially those involving relational databases with multi-table structures, matching often requires mapping multiple columns from the source schema to multiple columns in the target schema. We refer to this as an n:m mapping. This pattern, while expansive, ensures that important relationships are captured, even if it occasionally results in higher false positives. For most industrial applications, prioritizing recall (minimizing missed matches) is more desirable than optimizing precision, as it reduces the risk of missed correspondences~\cite{parciak2024schema}.


\begin{definition}
\vspace{-2mm}
\label{def:complexmatching}
In an n:m match, the task is to find a mapping \(\Psi:A_s \to A_t\), where each column \( a \in A_s \) is associated with a subset \(\Psi(a) \subseteq A_t\).

\end{definition}

%% file: sections/methodology.tex
\paragraph{Framework Design} Unifying different schema matching methods presents significant challenges due to the diversity in their design:
\begin{itemize}[label=•]
    \item \textbf{Task Scope Differences:} Some methods target table-to-table matching, aligning columns between isolated tables, while others tackle schema-to-schema matching, which captures relationships across multiple interrelated tables. These differing scopes highlight the need for a flexible framework that supports both fine-grained and holistic matching strategies.
    
    \item \textbf{Integrating Large Language Models (LLMs):} LLMs are becoming integral to schema matching, but their varying integration across methods raises fairness concerns. Some approaches leverage LLMs extensively, while others use them minimally or not at all. A fair framework should standardize LLM usage to facilitate meaningful comparisons between different approaches.
\end{itemize}

\begin{wrapfigure}{l}{0.53\textwidth}
\vspace{-5mm}
\begin{minipage}{\linewidth}
\scriptsize
\vspace{-4mm}
\input{figures/algorithm} 
\vspace{-2mm}
\captionof{algorithm}{\small \framework}
\label{alg:llmatch}
\end{minipage}
\vspace{-11mm}
\end{wrapfigure}

To address these challenges, we propose a \framework with a focus on structured evaluation and fair comparison. The processing framework includes three major steps: \emph{Schema Preparation}, \emph{Table Selection} and \emph{Column Matching}, as outlined in Algorithm~\ref{alg:llmatch}.

\noindent\textbf{Schema Preparation with Rollup} In this step, we organize schema information into three categories: table/column names (syntactic content), primary and foreign keys (structural relationships), and table/column descriptions (semantic metadata). This enables \framework to assess feature-level contribution to schema matching performance. 
To further reduce schema complexity and improve alignment performance, we introduce \emph{Rollup} (line 1), a transformation technique that simplifies both source and target schemas by merging semantically related columns into a single, higher-level abstraction. Unlike traditional rule-based processing, \emph{Rollup} relies on LLM to infer semantically related columns and determine meaningful merged aliases. For example, in Fig.~\ref{fig:rollup-drilldown}, \emph{Rollup} groups \texttt{currency},\texttt{currency\_code\_id} and \texttt{currency\_code} into a new alias \texttt{currency\_meta\_data}. This abstraction reduces the number of elements involved in matching and directs focus to broader semantic patterns rather than low-level features.
\emph{Rollup} is particularly useful in schemas where related information is distributed across multiple columns. Importantly, all rolled-up columns are recorded so that, after a high-level match is established, the original columns can be reintroduced in the \emph{Drilldown} phase for fine-grained alignment.

\vspace{2mm}
\noindent\textbf{Table Selection} This step (line 2) narrows the set of target tables for each source table \( T_s \), ensuring that subsequent matching focuses only on relevant candidates. We evaluate four strategies for candidate selection: (1) None: matching all target tables simultaneously without filtering, (2)Nested Join: processing each target table separately using isolated prompts, (3)Vector Similarity: selecting the top-\( k \) most similar tables based on embedding similarity, and (4)LLM: allowing the LLM to identify relevant tables using schema context. Each strategy reflects a trade-off between context size, semantic filtering, and computational efficiency.

\vspace{2mm}
\noindent\textbf{Column Matching with Drilldown} After relevant target tables are selected, column matching (line 3) is performed. To enhance alignment precision, we introduce \emph{Drilldown}, a refinement step that revisits the original components of previously rolled-up columns. As shown in Fig.~\ref{fig:rollup-drilldown}, once \texttt{source.bal\_currency} is matched to \texttt{target.currency\_meta\_data}, the \emph{Drilldown}  (line 7) phase triggers a second, focused matching process where the LLM reconsiders the original target columns—such as \texttt{target.currency}, \texttt{target.currency\_code}, and \texttt{target.currency\_id} as candidates for alignment. Unlike a simple expansion, \emph{Drilldown} does not assume that all rolled-up columns are relevant; instead, the LLM performs a more targeted evaluation within this reduced context. In some cases, only a subset of the rolled-up columns are aligned, while others may be excluded entirely. This two-step process balances abstraction and detail, enabling high-level semantic alignment followed by selective refinement.

%% file: figures/algorithm.tex
\scriptsize
\textbf{Inputs:} source schema $S_s = (T_s, A_s, D_s, R_s)$ and target schema $S_t = (T_t, A_t, D_t, R_t)$; LLM $\mathcal{F}$; embedding model $\Phi$; context limit $L$. \\
\textbf{Output:} A mapping $\Psi : A_s \to \mathcal{P}(A_t)$
\begin{algorithmic}[1]
\State $S_s^{roll}, S_t^{roll} \gets$ \textsc{Rollup}$(S_s, S_t, \mathcal{F})$
\State $T_s^{sel}, T_t^{sel} \gets$ \textsc{TableSelection}$(S_s^{roll}, S_t^{roll}, \mathcal{F}, L)$
\State $\Psi_{coarse} \gets$ \textsc{LLMColumnMatch}$(T_s^{sel}, T_t^{sel}, \mathcal{F})$
\State $\Psi \gets \emptyset$ 
\ForAll{$(a, b) \in \Psi_{coarse}$}
    \If{$b$ is a rolled-up alias}
        \State $C \gets$ \textsc{Drilldown}$(a, b, \mathcal{F})$
        \State $\Psi \gets \Psi \cup \{(a, c) \mid c \in C\}$
    \Else
        \State $\Psi \gets \Psi \cup \{(a, b)\}$
    \EndIf
\EndFor
\State \Return $\Psi$
\end{algorithmic}

%% file: sections/dataset.tex
\noindent Existing schema matching benchmarks often fail to capture the complexity of real-world industry scenarios. They typically consist of a single source and target table, with limited structural depth and no inter-table relationships. To address this gap, we collaborated with major financial institutions to design representative schema matching tasks. While confidentiality constraints prevent releasing of internal datasets, we propose \benchmark by curating public schema pairs that reflect the structural and semantic complexity of enterprise schemas. Drawn from healthcare, finance, and entertainment domains, \benchmark provides a foundation for evaluating schema matching in real-world applications.

\paragraph{Dataset Statistics} Table~\ref{tab:groundtruth} compares traditional benchmarks~\cite{koutras2021valentine} with \benchmark. \benchmark contains multiple tables and rich PK/FK relationships compared to flat single-table structures of existing datasets. \benchmark has higher mapping complexity with more frequent multiple-multiple table mappings and less 1:1 correspondences.

\paragraph{Mapping Development} \benchmark includes seven source–target schema pairs. \textit{Mimic-omop} and \textit{synthea-omop} are adapted from prior work~\cite{rematch,huang2150transform}; \textit{cms-omop}, \textit{cprd\_aurum-omop}, \textit{cprd\_gold-omop} are reverse-engineered from ETL code and verified by data scientists. The finance pair (\textit{bank1-bank2}) was anonymized and approved for release with industry partners. The entertainment pair (\textit{imdb-sakila}) was manually annotated by experts. All schemas and ground truth mappings are released at~\cite{sourcecode}.

\begin{table*}[th]
\vspace{-8mm}
\centering
\captionsetup{ justification=raggedright,singlelinecheck=false, labelsep=colon, ,width=\textwidth} 
\caption{{\small Dataset statistics for traditional benchmarks (top) and the complex benchmark (bottom) we developed. Complex datasets have more mapped target tables per source table and a low ratio of simple (1:1) mappings.}}

\include{tables/ground_truth_statistics}

\label{tab:groundtruth}
\vspace{-7mm}
\end{table*}

%% file: tables/ground_truth_statistics.tex
\scriptsize
\setlength{\tabcolsep}{3pt}
\renewcommand{\arraystretch}{0.85}
\begin{threeparttable}
  \begin{tabularx}{\textwidth}{@{}l c *{3}{>{\centering\arraybackslash}X}@{}}
    \toprule
    Task            & Domain  & Source Stats\tnote{a} & Target Stats\tnote{a} & Mappings Stats\tnote{b} \\
    \midrule
    mjs-mjt         & Entertainment    & 1/13/0/0            & 1/13/0/0            & 6 / 100\%/ 1/1 \\
    msjs-msjt       & Entertainment    & 1/14/0/0            & 1/14/0/0            & 8 / 100\%/ 1/1 \\
    mus-mut         & Entertainment    & 1/20/0/0            & 1/20/0/0            & 20 / 100\%/ 1/1 \\
    mvs-mvt         & Entertainment    & 1/13/0/0            & 1/13/0/0            & 6 / 100\%/ 1/1 \\
    \midrule
    imdb-sakila     & Entertainment    & 7/39/2/7            & 16/90/12/22         & 22 / 26.7\%/ 1.0/3 \\
    bank1-bank2     & Finance    & 9/27/5/11           & 9/36/4/12           & 11 / 100\%/ 0.4/1 \\
    cms-omop        & Healthcare  & 5/96/1/4            & 39/432/12/58        & 157 / 3.7\%/ 7.4/11 \\
    synthea-omop    & Healthcare  & 12/111/3/19         & 39/432/12/58        & 101 / 19.4\%/ 1.5/3 \\
    cprd\_aurum-omop& Healthcare  & 8/76/5/21           & 39/432/12/58        & 42 / 48.0\%/ 1.8/3 \\
    cprd\_gold-omop & Healthcare  & 9/123/4/21          & 39/432/12/58        & 52 / 25.0\%/ 2.4/4 \\
    mimic\_iii-omop & Healthcare  & 26/324/6/55         & 39/432/12/58        & 189 / 14.7\%/ 1.5/5 \\
    \bottomrule
  \end{tabularx}
  \begin{tablenotes}[flushleft]
    \item[a] Number of tables/columns/primary keys/foreign keys.
    \item[b] Total number of mappings/percentage of 1:1 mappings/ average number of target tables per source table/max number of target tables per source table.
  \end{tablenotes}
\end{threeparttable}

%% file: sections/evaluation.tex
We evaluate the following baselines using our proposed \framework framework.
\begin{itemize}[label=•]
\item \coma~\cite{do2002coma}: A classical schema matching approach that combines multiple schema-based matchers, representing schemata as rooted graphs.

\item \similarityflooding~\cite{melnik2002similarity}: Transforms schemas into directed graphs and propagates similarity scores through neighboring nodes.

\item \cupid~\cite{cupid}: Represents schemas as hierarchical tree structures and calculates similarity as a weighted sum of linguistic (name-based) and structural (context-based) similarities. 

\item \unicorn~\cite{unicorn}: A the-state-of-the-art deep learning model that learns from multiple datasets and tasks using a mixture-of-experts model. 

\item \rematch~\cite{rematch}: An LLM-based method that serializes tables into documents, generates document embeddings, and uses vector similarity to pair the most similar target tables with the source table. The selected tables are then fed into an LLM to produce column mappings.

\item \framework (our approach): We leverage LLMs for both table candidate selection and column matching. Similar to \rematch, we include schema information such as names, primary/foreign keys, and descriptions in the prompt. Details of the two prompt templates and schema serialization format are provided in the extended report \cite{sourcecode}. 
\end{itemize}

\begin{table*}[th]
\vspace{-5mm}
\centering
\include{tables/full_experiment_f1_score}
\captionsetup{ justification=raggedright,singlelinecheck=false, width=\textwidth}
\caption{\small F1 scores of schema matching methods on simple and complex tasks. Traditional methods (\coma, \similarityflooding, \cupid, \unicorn) perform well on simple tasks but degrade significantly on complex ones. In contrast, LLM-based approaches (\rematch and \framework) maintain high performance across both, highlighting their effectiveness on complex schema matching.}
\label{table:all_result_end_to_end}
\vspace{-9mm}
\end{table*}

\subsection{Evaluation Metric}
We use \textbf{F1 score} as our primary metric, following prior schema matching works~\cite{parciak2024schema,zhang2021smat,alwan2017survey,do2002coma}. While alternatives like recall@k~\cite{unicorn} and accuracy@k~\cite{rematch} exist, we prioritize F1 as our method treats each match equally and does not produce ranked outputs. During evaluation, we treat foreign key (FK) matches as equal to their corresponding primary keys (PKs).

\subsection{Implementation} We use \texttt{gpt-3.5-turbo} and \texttt{gpt-4o-mini} models with default settings, without any parameter modifications or fine-tuning. As LLM outputs are inherently non-deterministic~\cite{openai2023seed}, we acknowledge the limited reproducibility of specific responses. All embeddings are generated using SBERT~\cite{reimers-2019-sentence-bert}. Since prompt phrasing can significantly affect LLM behavior~\cite{schulhoff2024prompt}, we release all prompt templates and schema serialization formats in the full report~\cite{sourcecode}.

\subsection{Evaluation Goals} We evaluate the effectiveness of \framework by addressing following questions: how different approaches perform on traditional and new benchmarks (Sec.~\ref{sec:overall_evalation}), the impact of table selection methods on performance (Sec.~\ref{sec:table_selection_study}), and the role of schema elements such as names, descriptions, and PK/FK relationships (Sec.~\ref{sec:schema_ablation_study}). Additionally, we examine how multi-level schema matching with Rollup and Drilldown affects performance (Sec.~\ref{sec:effect_of_rollup_and_drilldown}), the implications of LLM context limits for large datasets (Sec.~\ref{sec:scalability_study}), and the productivity gains in schema matching with and without machine assistance (Sec.~\ref{sec:productivity_gain_study}).

\subsection{End-to-End Performance Evaluation}
\label{sec:overall_evalation}
The performance of \framework are shown in Table~\ref{table:all_result_end_to_end}. For simple tasks, both traditional and LLM-based methods perform well, with high scores across all datasets. However, traditional methods degrade sharply on complex tasks, consistent with prior findings~\cite{zhang2023schema}. On complex benchmarks, \framework outperforms all baselines. For example, on the largest dataset, \textit{mimic\_iii-omop}, \framework achieves an F1 score of 0.4, double that of \rematch (0.2), and far above \coma (0.04) and \similarityflooding (0.09).One exception is \textit{bank1-bank2}, where traditional methods perform comparably due to extensive shared vocabulary in the schema, which is effectively picked up by lexical matchers. Fig.~\ref{fig:performance_increase} shows the performance gap between \framework and \rematch plotted against task complexity. As tasks become more challenging, \framework's advantage grows, especially with GPT-3.5-turbo, highlighting the method's robustness even with less capable models.

\subsection{Table Selection Strategy Study}
\label{sec:table_selection_study}
\begin{table*}
\centering
\vspace{-4mm}

\include{tables/table_selection_strategy_combined}
\captionsetup{ justification=raggedright,singlelinecheck=false, labelsep=colon,}
\caption{\small Results are F1 scores for various table selection methods, with the number of selected tables in parentheses. LLM-based methods select fewer tables but achieve higher quality.}
\vspace{-9mm}

\label{table:table_selection_results}

\end{table*}

To evaluate the impact of different table selection strategies (Step 2 in Fig.~\ref{fig:flowchart}), we compare four methods: \textit{None} includes all target tables for each source table in a single prompt. \textit{Nested Join} processes each source–target pair independently using separate prompts. \textit{Vector Similarity} selects the top-$k$ most similar target tables using embedding-based scores. \textit{LLM} prompts the model with source and target table descriptions, asking it to identify relevant candidates directly. Table~\ref{table:table_selection_results} reports the F1 score and the average number of target tables selected per source table. 
\textit{LLM} consistently outperforms \textit{Vector Similarity}, suggesting that while \textit{Vector Similarity} is often the default option in semantic retrieval tasks, it may not be the best option for schema matching.

\begin{figure}[h]
\vspace{-8mm}
\centering
\begin{minipage}[t]{0.45\linewidth}
    \centering
    \include{figures/performance_gain_per_dataset}
    \vspace{-11mm}
    \captionof{figure}{\small Percentage improvement of \framework\ over the SOTA baseline (\rematch) across datasets of varying complexity.}
    \label{fig:performance_increase}
\end{minipage}%
\hspace{4mm}
\begin{minipage}[t]{0.48\linewidth}
    \centering
    \include{figures/schema_element_study}
    \vspace{-11mm}
    \captionof{figure}{\small Schema element ablation study: adding contextual information such as PK/FK and descriptions improves the F1 score.}
    \label{fig:schema_element_study}
\end{minipage}
\vspace{-10mm}
\end{figure}

\subsection{Schema Elements Ablation Study}
\label{sec:schema_ablation_study}
To evaluate the contribution of schema elements to matching performance, we conducted an ablation study. Figure~\ref{fig:schema_element_study} shows F1 scores across different element combinations. Results indicate that adding \textit{description} and \textit{relationship} improves performance, with the full combination achieving the highest scores.

\subsection{Effect of Rollup and Drilldown}
\label{sec:effect_of_rollup_and_drilldown}
Fig.~\ref{fig:relative_improvement} illustrates the impact of Rollup and Drilldown across different datasets. The performance gains vary with schema complexity. Datasets with more fragmented or relational structures, such as \textit{cprd\_gold-omop}, \textit{mimic\_iii-omop}, and \textit{synthea-omop}, show the largest improvements. In contrast, simpler datasets like \textit{imdb-sakila} and \textit{bank1-bank2} exhibit only modest gains. These results suggest that the benefits of Rollup and Drilldown are most pronounced in complex matching scenarios, where hierarchical alignment is critical for resolving multi-level relationships.

\subsection{Scalability Study}\label{sec:scalability_study}
To assess scalability, we constrain input context size while leaving output unrestricted. For consistency across LLMs, we approximate context size by word count. Tasks exceeding a single-prompt limit are split into multiple smaller prompts, ensuring the largest source and target tables fit together. This constraint applies to both table selection and column matching. As shown in Fig.~\ref{fig:scalability_study}, overly small contexts fragment the task and degrade performance. These results highlight the importance of sufficient context capacity for effective schema matching on large, complex datasets.

\begin{figure}[h]
    \centering

    \begin{minipage}[t]{0.52\linewidth}
        \centering
        \vspace{-3mm}
        \include{figures/effect_of_rollup_drilldow}
        \vspace{-10mm}
        \captionof{figure}{\small Rollup/Drilldown improves matching accuracy, with gains up to 40\% on some datasets.}
        \label{fig:relative_improvement}
        \vspace{-4mm}\include{figures/productivity_gain_study_horizontal}
           \vspace{-10mm}
        \captionof{figure}{\small Productivity Gain Study. Machine assistance significantly improves matching F1 score with equal time per task.}
        \label{fig:productivity_gain_study}
    \end{minipage}
    \hfill
    \begin{minipage}[t]{0.45\linewidth}
        \centering
        \include{figures/context_size_study}
           \vspace{-12mm}
        \captionof{figure}{\small Scalability Study. Performance drops with small context windows and plateaus once the schema fits entirely.}
        \label{fig:scalability_study}
    \end{minipage}
    \vspace{-11mm}
\end{figure}

\subsection{Productivity Gain Study}
\label{sec:productivity_gain_study}
To evaluate the impact of LLM-generated matches on productivity, we asked annotators to manually align two schemas under two conditions: with and without machine-generated assistance. Each task was limited to two minutes. As shown in Fig.~\ref{fig:productivity_gain_study}, the average F1 across four participants was notably higher with machine assistance, demonstrating the helpfulness of LLM outputs. These findings align with findings from~\cite{zhang2023schema}, who reported that LLM can reduce labeling costs by up to 81\% compared to manual methods.

%% file: tables/full_experiment_f1_score.tex
\centering
\scriptsize
\setlength{\tabcolsep}{3pt}
\renewcommand{\arraystretch}{0.85}
\begin{threeparttable}
  \begin{tabular}{l|c c c c c c c|c c c c}
    \toprule
    & \multicolumn{7}{c|}{\textbf{Complex Matching Task}} & \multicolumn{4}{c}{\textbf{Simple Matching Task}} \\
    \cmidrule(lr){2-8} \cmidrule(lr){9-12}
    \textbf{Strategy} & 1 & 2 & 3 & 4 & 5 & 6 & 7 & 8 & 9 & 10 & 11 \\
    \midrule
    \coma                    & \underline{0.74} & 0.18 & 0.13 & 0.03 & 0.08 & 0.01 & 0.04 & 0.91 & 0.71 & 0.57 & \textbf{0.91} \\
    \similarityflooding     & 0.50 & 0.21 & 0.08 & 0.03 & 0.03 & 0.04 & 0.09 & 0.71 & 0.70 & 0.89 & 0.59 \\
    \cupid                   & 0.32 & 0.00 & 0.00 & 0.00 & 0.00 & 0.01 & 0.00 & 0.33 & 0.80 & 0.86 & 0.33 \\
    \unicorn                 & 0.38 & 0.00 & 0.05 & 0.01 & 0.03 & 0.05 & 0.01 & \textbf{0.92} & 0.80 & 0.87 & 0.80 \\
    \midrule
    \rematch~GPT-3.5        & 0.40 & 0.36 & 0.06 & 0.04 & 0.06 & 0.07 & 0.07 & 0.86 & 0.70 & 0.88 & 0.67 \\
    \rematch~GPT-4o-mini    & \underline{0.74} & \underline{0.64} & \underline{0.20} & \underline{0.19} & 0.19 & 0.16 & \underline{0.20} & \textbf{0.92} & \underline{0.84} & 0.81 & 0.71 \\
    \framework~GPT-3.5      & 0.67 & 0.47 & 0.15 & 0.16 & \underline{0.26} & \underline{0.20} & 0.13 & \textbf{0.92} & 0.82 & \underline{0.95} & \underline{0.83} \\
    \framework~GPT-4o-mini  & \textbf{0.85} & \textbf{0.73} & \textbf{0.36} & \textbf{0.30} & \textbf{0.37} & \textbf{0.37} & \textbf{0.33} & \textbf{0.92} & \textbf{0.94} & \textbf{0.97} & \underline{0.83} \\
    \bottomrule
  \end{tabular}
  \begin{tablenotes}[flushleft]
    \item Columns 1 bank1-bank2; 2 imdb-sakila; 3 cprd\_au-omop; 4 cprd\_gold-omop; 5 synthea-omop; 6 cms-omop; 7 mimic\_iii-omop; 8 mjs-mjt; 9 msjs-msjt; 10 mus-mut; 11 mvs-mvt. Columns 1–7 correspond to Complex Schema Matching tasks, Columns 8–11 to Simple Schema Matching tasks.
  \end{tablenotes}
\end{threeparttable}

%% file: tables/table_selection_strategy_combined.tex
\scriptsize
\setlength{\tabcolsep}{3pt}
\begin{threeparttable}
  \newcommand{\scell}[2]{#1/\scriptsize(#2)}
  \begin{tabularx}{\textwidth}{l *{7}{>{\centering\arraybackslash}X}}
    \toprule
    \textbf{Strategy} & \textbf{1} & \textbf{2} & \textbf{3} & \textbf{4} & \textbf{5} & \textbf{6} & \textbf{7} \\
    \midrule
    None                & \scell{0.74}{9.0} & \scell{0.39}{16.0} & \scell{0.28}{39.0} & \scell{0.22}{39.0} & \scell{0.23}{39.0} & \scell{0.21}{39.0} & \scell{0.13}{39.0} \\
    Nested Join         & \scell{0.64}{1.0} & \scell{0.32}{1.0} & \scell{0.20}{1.0} & \scell{0.18}{1.0} & \scell{0.10}{1.0} & \scell{0.10}{1.0} & \scell{0.07}{1.0} \\
    VS\tnote{a} (top5)  & \scell{\underline{0.85}}{5.0} & \scell{\underline{0.67}}{5.0} & \scell{0.19}{5.0} & \scell{0.27}{5.0} & \scell{0.29}{5.0} & \scell{0.35}{5.0} & \scell{0.16}{5.0} \\
    VS\tnote{a} (top10) & \scell{\textbf{0.88}}{9.0} & \scell{\underline{0.67}}{10.0} & \scell{\underline{0.33}}{10.0} & \scell{\textbf{0.37}}{10.0} & \scell{0.29}{10.0} & \scell{\underline{0.36}}{10.0} & \scell{0.25}{10.0} \\
    VS\tnote{a} (top15) & \scell{\textbf{0.88}}{9.0} & \scell{0.59}{15.0} & \scell{0.27}{15.0} & \scell{0.29}{15.0} & \scell{\underline{0.32}}{15.0} & \scell{0.31}{15.0} & \scell{\underline{0.26}}{15.0} \\
    LLM\tnote{b}        & \scell{\underline{0.85}}{4.6} & \scell{\textbf{0.73}}{4.0} & \scell{\textbf{0.36}}{6.3} & \scell{\underline{0.30}}{6.9} & \scell{\textbf{0.37}}{4.7} & \scell{\textbf{0.37}}{5.4} & \scell{\textbf{0.33}}{6.9} \\
    \bottomrule
  \end{tabularx}
  \begin{tablenotes}[flushleft]
    \item[1] Column order: 1 bank1-bank2; 2 imdb-sakila; 3 cprd\_aurum-omop; 4 cprd\_gold-omop; 5 synthea-omop; 6 cms-omop; 7 mimic\_iii-omop. [a] VS: \texttt{Vector Similarity} [b] LLM: gpt-4o-mini
  \end{tablenotes}
\end{threeparttable}

%% file: figures/performance_gain_per_dataset.tex
\begin{tikzpicture}
\begin{axis}[
    width=\linewidth,
    height=0.8\linewidth,
    xlabel={\shortstack{Average Number of Target Tables\\per Source Table}},
    ylabel={Performance Increase (\%)},
    xmin=0, xmax=8,
    ymin=0, ymax=550,
    xtick={0,1,2,3,4,5,6,7,8},
    ytick={0,100,200,300,400,500},
    legend style={
        at={(0.5,1.2)},
        anchor=north,
        font=\bfseries\scriptsize,
        legend columns=2,
        /tikz/every even column/.append style={column sep=0.2cm},
        draw=none,fill=none
    },
    grid=both,
    grid style={line width=.1pt, draw=gray!20},
    major grid style={line width=.2pt, draw=gray!50},
    minor tick num=1,
    xlabel style={font=\small},
    ylabel style={font=\small},
    label style={font=\small},
    ticklabel style={font=\small},
]

\addplot[
    color=red!80,
    mark=*,
    thick,
]
coordinates {
    (0.44, 42.50)    
    (1.00, 64.71)    
    (1.50, 280.00)   
    (1.54, 142.86)   
    (1.75, 300.00)   
    (2.44, 525.00)   
    (7.40, 350.00)   
};
\addlegendentry{\texttt{GPT-3.5-turbo}}

\addplot[
    color=blue!80,
    mark=square*,
    thick,
]
coordinates {
    (0.44, 1.35)     
    (1.00, 14.06)    
    (1.50, 94.74)    
    (1.54, 65.00)    
    (1.75, 80.00)    
    (2.44, 57.89)    
    (7.40, 131.25)   
};
\addlegendentry{\texttt{GPT-4o-mini}}

\addplot[
    only marks,
    mark size=2.5pt,
    color=red!80,
    nodes near coords,
    nodes near coords style={font=\scriptsize, anchor=north east},  
    point meta=explicit symbolic,
]
coordinates {
    (0.44, 67.50)    
    (1.00, 30.56)    
    (1.50, 333.33)   
    (1.54, 85.71)    
    (1.75, 150.00)   
    (2.44, 300.00)   
    (7.40, 185.71)   
};

\end{axis}
\end{tikzpicture}

%% file: figures/schema_element_study.tex
\pgfplotslegendfromname{combinedlegend}
\begin{tikzpicture}
    \begin{groupplot}[
        group style={
            group size=1 by 2,
            vertical sep=0.1cm,
        },
        width=\linewidth,
        height=0.5\linewidth,
        xtick={1,2,3,4,5,6,7},
        xticklabels={
            bank1-bank2,
            imdb-sakila,
            cprd\_aurum-omop,
            cprd\_gold-omop,
            synthea-omop,
            cms-omop,
            mimic\_iii-omop
        },
        x tick label style={rotate=18, anchor=east, xshift=30pt, yshift=-4pt, font=\scriptsize},
        y tick label style={font=\scriptsize},
        ylabel style={yshift=-3.5ex, font=\scriptsize},
        ymajorgrids=true,
        grid style={solid, gray},
        legend to name=combinedlegend,
        legend style={
            legend columns=2,
            font=\scriptsize,
            /tikz/every even column/.append style={column sep=0cm},
            draw=none,
            legend cell align=left,
            fill=none
        },
    ]

    \nextgroupplot[
        title={GPT-3.5-Turbo},
        title style={yshift=-5ex, font=\scriptsize},
        ylabel={F1 Score},
        ymin=0, ymax=0.9,
        xtick=\empty,
        legend to name=combinedlegend,
        legend style={draw=none},
    ]

    \addplot[color=blue, mark=*, solid, line width=1pt] coordinates {
        (1,0.5714) (2,0.6667) (3,0.2692) (4,0.1975) (5,0.2695) (6,0.2513) (7,0.2164)
    };
    \addlegendentry{LLM (GPT-3.5-Turbo)}

    \addplot[color=red, mark=square*, solid, line width=1pt] coordinates {
        (1,0.6667) (2,0.6486) (3,0.1795) (4,0.15) (5,0.25) (6,0.2028) (7,0.1711)
    };
    \addlegendentry{LLM-No Desc No FK (GPT-3.5-Turbo)}

    \addplot[color=green, mark=triangle*, solid, line width=1pt] coordinates {
        (1,0.7) (2,0.6111) (3,0.161) (4,0.1446) (5,0.2937) (6,0.263) (7,0.1858)
    };
    \addlegendentry{LLM-No Description (GPT-3.5-Turbo)}

    \addplot[color=orange, mark=diamond*, solid, line width=1pt] coordinates {
        (1,0.6923) (2,0.6486) (3,0.1356) (4,0.1892) (5,0.2737) (6,0.2460) (7,0.1818)
    };
    \addlegendentry{LLM-No FK (GPT-3.5-Turbo)}

    \nextgroupplot[
        title={GPT-4o-mini},
        title style={yshift=-5ex, font=\scriptsize},
        ymin=0, ymax=0.9,
        legend to name=combinedlegend,
        legend style={draw=none},
        ylabel={F1 Score},
    ]

    \addplot[color=blue, mark=o, solid, line width=1pt] coordinates {
        (1,0.8) (2,0.6857) (3,0.3529) (4,0.4301) (5,0.4231) (6,0.3128) (7,0.4022)
    };
    \addlegendentry{name+desc.+PK/FK}

    \addplot[color=red, mark=square, solid, line width=1pt] coordinates {
        (1,0.8333) (2,0.6486) (3,0.25) (4,0.2316) (5,0.3627) (6,0.2896) (7,0.2398)
    };
    \addlegendentry{name}

    \addplot[color=green, mark=triangle, solid, line width=1pt] coordinates {
        (1,0.6) (2,0.5641) (3,0.2571) (4,0.2947) (5,0.3592) (6,0.2280) (7,0.2616)
    };
    \addlegendentry{name+PK/FK}

    \addplot[color=orange, mark=diamond, solid, line width=1pt] coordinates {
        (1,0.6667) (2,0.6471) (3,0.2989) (4,0.2889) (5,0.3645) (6,0.3203) (7,0.3306)
    };
    \addlegendentry{name+desc.}

    \end{groupplot}
\end{tikzpicture}

%% file: figures/effect_of_rollup_drilldow.tex
\begin{tikzpicture}
\begin{axis}[
    width=0.9\linewidth,
    height=0.5\linewidth,
    ylabel={F1 Score},
    xlabel={},
    xtick=data,
    xticklabels={
        bank1-bank2,
        imdb-sakila,
        cprd\_aurum-omop,
        cprd\_gold-omop,
        synthea-omop,
        cms-omop,
        mimic\_iii-omop
    },
    x tick label style={rotate=20,, xshift=25pt, yshift=-10pt, anchor=east, font=\scriptsize},
    yticklabel style={font=\scriptsize, },
    legend style={font=\scriptsize, at={(0.5,0.9)}, anchor=south, legend columns=2, draw=none,fill=none},
    grid=both
]

\addplot[
    color=red,
    mark=*,
    thick
] coordinates {
    (0, 0.85)
    (1, 0.73)
    (2, 0.36)
    (3, 0.30)
    (4, 0.37)
    (5, 0.37)
    (6, 0.33)
};
\addlegendentry{with}

\addplot[
    color=blue,
    mark=square*,
    thick
] coordinates {
    (0, 0.8000)
    (1, 0.7095)
    (2, 0.3302)
    (3, 0.2001)
    (4, 0.3386)
    (5, 0.3526)
    (6, 0.2630)
};
\addlegendentry{without Rollup/Drilldown}

\end{axis}
\end{tikzpicture}

%% file: figures/productivity_gain_study_horizontal.tex
\begin{tikzpicture}
\begin{axis}[
    xbar,
    xmin=0,
    xmax=0.8,
    width=0.75\linewidth,
    height=0.4\linewidth,
    bar width=5pt,
    title={\scriptsize{F1 Score}}, 
    title style={yshift=-8pt},     
    symbolic y coords={With Assist, Without Assist},
    ytick=data,
    xtick=\empty,
    nodes near coords,
    nodes near coords align={horizontal},
    enlarge y limits=0.3,
    ticklabel style={font=\scriptsize},
    ytick style={draw=none},
    xtick style={draw=none},
]
\addplot coordinates {(0.5813,With Assist) (0.1377,Without Assist)};
\end{axis}
\end{tikzpicture}

%% file: figures/context_size_study.tex
\begin{tikzpicture}
    \begin{axis}[
        width=\linewidth,
        height=0.71\linewidth,
        xlabel={LLM Context Size},
        ylabel={F1 Score},
        xmin=1000, xmax=10000,
        ymin=0, ymax=1,
        xtick={1000,2000,5000,10000},
        xticklabels={
            1k, 
            2k, 
            5k, 
            10k
        },
        ytick={0,0.2,0.4,0.6,0.8,1.0},
  legend style={
    at={(0.5,1.1)},       
    anchor=south,         
    font=\scriptsize,
    draw=none,fill=none,
    legend columns=2,     
    /tikz/every even column/.append style={column sep=0.01cm},
    legend cell align=left
},
        grid=both,
        grid style={line width=.1pt, draw=gray!20},
        major grid style={line width=.2pt, draw=gray!50},
        xlabel style={font=\small, yshift=3pt},
        ylabel style={font=\scriptsize},
        label style={font=\scriptsize},
        ticklabel style={font=\scriptsize},
        ylabel style={yshift=-3ex},
         scaled x ticks=false, 
    ]

    \addplot+[mark=*, color=blue, thick] coordinates {
        (1000, 0.2857) (2000, 0.5882) (5000, 0.5882) (10000, 0.6286)
    };
    \addlegendentry{imdb-sakila}
  \addplot+[mark=diamond*, color=purple, thick] coordinates {
        (1000, 0.1389) (2000, 0.25) (5000, 0.2198) (10000, 0.2299)
    };
    \addlegendentry{cprd\_gold-omop}
    \addplot+[mark=square*, color=red, thick] coordinates {
        (1000, 0.0339) (2000, 0.2121) (5000, 0.2222) (10000, 0.2679)
    };
    \addlegendentry{cms-omop}

    \addplot+[mark=triangle*, color=green, thick] coordinates {
        (1000, 0.22) (2000, 0.1872) (5000, 0.2667) (10000, 0.3333)
    };
    \addlegendentry{cprd\_aurum-omop} 

    \addplot+[mark=o, color=cyan, thick] coordinates {
        (1000, 0.1605) (2000, 0.2543) (5000, 0.3109) (10000, 0.2328)
    };
    \addlegendentry{synthea-omop}

    \addplot+[mark=x, color=orange, thick] coordinates {
        (1000, 0.0539) (2000, 0.1101) (5000, 0.2442) (10000, 0.1754)
    };
    \addlegendentry{mimic\_iii-omop}
    
    \addplot+[mark=star, color=black, thick] coordinates {
        (1000, 0.81) (2000, 0.83) (5000, 0.83) (10000, 0.8)
    };
    \addlegendentry{bank1-bank2}
    \end{axis}
\end{tikzpicture}

%% file: sections/conclusion.tex
This paper presents a framework that decomposes schema matching into three stages: schema preparation, table selection, and column matching. Our key contribution is the introduction of Rollup and Drilldown, a multi-level schema matching addon that simplifies complex schemas prior to matching and refines alignments afterward. Using this framework, we also observed that vector similarity, commonly employed for table selection, is not the most effective method, offering valuable implications for other LLM-based applications. Furthermore, we developed and publicly released a comprehensive, multi-industry benchmark for complex schema matching to support future research in this domain.